\documentclass[preprint,preprintnumbers,showpacs,amsmath,amssymb,aps,prb]{revtex4-1}
\usepackage{graphicx}
\usepackage{dcolumn}
\usepackage{bm}

\begin{document}

\title{Magnetic levitation of metamaterial bodies enhanced with magnetostatic surface resonances}

\author{Yaroslav Urzhumov$^{1}$, Wenchen Chen$^{2}$, Chris Bingham$^{2}$, Willie Padilla$^{2}$ and David R. Smith$^{1}$}


\affiliation{\vspace{5 pt}$^1$Center for Metamaterials and Integrated Plasmonics,\\
Duke University, Durham, N.~C., 27708 USA\\
\vspace{5 pt}
$^2$Department of Physics, Boston College, Chestnut Hill, Mass. 02467 USA
}

\newcommand{\ba}{\begin{eqnarray}}
\newcommand{\ea}{\end{eqnarray}}
\newcommand{\be}{\begin{equation}}
\newcommand{\ee}{\end{equation}}
\newcommand{\para}{\parallel}

\def \d{\partial}
\def \Re{{\rm Re}}
\def \Im{{\rm Im}}
\def \diag{{\rm diag}}
\def \const{{\rm const}}
\def \eff{{\rm eff}}
\def \sign{{\rm sign}}

\begin{abstract}
We propose that macroscopic objects built from negative-permeability metamaterials may experience resonantly enhanced magnetic force in low-frequency magnetic fields. Resonant enhancement of the time-averaged force originates from magnetostatic surface resonances (MSR) which are analogous to the electrostatic resonances of negative-permittivity particles, well known as surface plasmon resonances in optics. We generalize the classical problem of MSR of a homogeneous object to include anisotropic metamaterials, and consider the most extreme case of anisotropy where the permeability is negative in one direction but positive in the others. It is shown that deeply subwavelength objects made of such indefinite (hyperbolic) media exhibit a pronounced magnetic dipole resonance that couples strongly to uniform or weakly inhomogeneous magnetic field and provides strong enhancement of the magnetic force, enabling applications such as enhanced magnetic levitation.
\end{abstract}


\maketitle

\section{\label{sec:intro} Introduction and motivation}

Electromagnetic (EM) forces are used in a variety of technologies, ranging from motors, EM guns~\cite{kitzmiller_murphy97} and magnetic levitation trains~\cite{schultz_funk05} to micro- and nano-actuators~\cite{shinozawa_kondo97} and optical tweezers~\cite{grigorenko_zhang08,wang_crozier11}. The spatial scales where EM forces are important to mankind range from atomic to cosmic~\cite{parker79}, and the relevant EM frequencies cover the entire spectrum from statics to gamma-ray frequencies.

For applications dealing with human scales ($\sim 1$~m),
high-intensity EM fields necessary for the generation of strong EM forces interfere with other devices and may be harmful to biological tissues.
The severity of this problem is substantially reduced if the fields are predominantly magnetic, since virtually all biological substances and the majority of conventional materials are almost purely dielectric and thus transparent to magnetic fields. While at a finite frequency it is impossible to completely eliminate the electric fields from the picture, the ratio $E/H$ can be strongly suppressed in the near field zone of a magnetic dipole oscillator, such as a conductor loop carrying alternating current (AC). For that reason, a large variety of EM force applications, including EM motors, maglev trains and magnetic brakes, have a strong preference for magnetic rather than electric forces.

Natural media with strong magnetic response, although readily available, provide only a limited range of magnetic permeability ($\mu$) values;
negative permeability is particularly hard to find. Negative $\mu$ can be found in magnetized ferrites in narrow frequency bands associated with ferromagnetic~\cite{he_vittoria06} and gyromagnetic~\cite{thompson55} resonances. However, magnetic permeability typically has a large loss tangent in those bands, leading to high specific absorption rates. As we show in this paper, negative $\mu$ with low enough loss tangent enables resonant enhancement of the magnetic polarizability; therefore, it is important to find or construct low-density media with this property.

Artificial media such as metamaterials have been demonstrated to provide negative $\mu$ with relatively low loss tangents~\cite{pendry_stewart99,smith_schultz00,smith_science01}.
Magnetic metamaterials consisting of metallic lines on thin dielectric substrates can be readily and inexpensively fabricated using the existing printed circuit board technologies (PCB). PCB metamaterials consisting of thin metal lines on low-density dielectric substrates have a substantial volume fraction of air and thus typically have much smaller density (of order 0.1~g/cm$^3$) than ferrites and other naturally magnetic media.
Their loss tangents can be reduced by using low-loss dielectric substrates and high-conductivity metals.
Additionally, the highly porous structure of PCB metamaterials allows for convection cooling, enabling higher peak intensities that do not cause material damage, which is impossible with homogeneous natural magnetics.
The operational band of negative-$\mu$ PCB metamaterials can be easily adjusted to any frequency between roughly 1~MHz to about 100~GHz, even without conceptual changes in the PCB fabrication process, by adjustments to the unit cell size, as well as inductance and capacitance of the resonating metamaterial elements.

Resonant enhancement of electric dipole polarizability by virtue of the so-called Surface Plasmon Resonance (SPR) is well-known in optics~\cite{shvets_urzh_jopa05,shvets_urzh_jopa06,lomakin_optexp06}. Its closely related analog, Surface Phonon Polariton Resonance (SPPR) can be found in mid-infrared~\cite{korobkin_josab06,korobkin_ap07,urzh_jopa07}.
Surface plasmon resonances are electrostatic in nature and can be understood as eigenmodes, that is, source-free solutions, of the electrostatic Laplace equation~\cite{shvets_urzh_prl04,shvets_urzh_jopa05,urzh_shvets_spie07}
\be
\nabla \epsilon \nabla \phi=0.
\label{eq:electrostatic}
\ee
When all materials involved have positive permittivity $\epsilon>0$, a uniqueness theorem can be proven that precludes the possibility of non-trivial solutions to this equation~\cite{bergman_stroud92,fredkin03}. However, when two or more homogeneous media with both negative and positive $\epsilon$ are included, source-free solutions exist at certain, negative values of the permittivity~\cite{fuchs75,bergman_stroud92,fredkin03}. In binary (two-medium) composites, these values depend exclusively on the shape of the positive-negative $\epsilon$ boundary~\cite{fuchs75,bergman_stroud92,fredkin03}, but not on the absolute size of the structure, since electrostatic problems do not have a spatial scale associated with the wavelength.

For simple geometric shapes, resonant values of $-\epsilon$ are not very large, of order unity: for a sphere, the dipole resonance occurs at $\epsilon_{res}=-2$, for a cylinder, it occurs at $\epsilon_{res}=-1$. The value of $-\epsilon_{res}$ can be increased by one or two orders of magnitude by introducing large aspect ratios and/or small gaps between the negative-$\epsilon$ particles. Still, relative permittivity in the range $-100$ to $-1$ is readily found in natural media only at optical (UV through mid-IR) frequencies, which explains why surface plasmon resonances have so far been a subject of mostly optical studies, and attracted little attention from the point of view of RF and microwave applications. On the other hand, metamaterials can easily deliver those moderately negative values of $\epsilon$ at any frequency of interest, including the microwave and lower-frequency regions.

Magnetostatic fields in the absence of electric currents can be described by the magnetostatic potential $\phi^{(m)}$, which satisfies the equation
$\nabla \mu \nabla \phi^{(m)}=0$, mathematically identical to the Laplace equation of electrostatics (\ref{eq:electrostatic}). This analogy is a particular case of EM duality, which can be stated as the exact equivalence of electric and magnetic field equations in the absence of free electric (and magnetic) charges. We can thus duplicate the electrostatic resonance phenomena using negative permeability media. This conclusion would also apply in the quasistatic, finite-frequency case, as long as the retardation parameter $D/\lambda$, where $D$ is the characteristic spatial scale of the problem, is negligibly small.

Existence of electric charges and their currents however creates a distinction between electrostatic and magnetostatic resonances.
While the negative-$\epsilon$ bands found in natural and artificial media can extend to arbitrarily low frequencies (recall the Drude dispersion of electron fluid), one cannot achieve negative $\mu(\omega)$ at zero frequency~\cite{wood_pendry07}, as the latter would imply negative energy density, which violates the conditions of thermodynamic stability. Negative permeability always occurs in finite-bandwidth portions of the spectrum surrounded by positive-$\mu$ bands; in that regard, magnetostatic surface resonances (MSR) studied in this paper resemble quasi-electrostatic surface resonances in plasmon-polaritonic structures~\cite{korobkin_josab06,korobkin_ap07,urzh_jopa07}. Within the already narrow negative-$\mu$ band, we predict a series of even narrower resonances associated with magnetic surface eigenmodes of the macroscopic magnetizable body.

In this paper, we consider metamaterial bodies that are much smaller than the free-space wavelength $\lambda$ of EM radiation at the operational frequency: $D\ll \lambda$. At the same time, we require that the unit cells of the metamaterial are much smaller than the whole body, $a\ll D$. This allows us to treat it as a homogeneous permeability object, and use metamaterial properties calculated from standard frequency-domain homogenization theories, both quasistatic~\cite{urzh_shvets_spie07} and full-wave~\cite{smith_prb02}. The only currently known method for achieving negative-$\mu$ in metamaterials is the embedding of self-resonant metamaterial resonators, which thus need to be much smaller than $D$. The hierarchy $a\ll D\ll \lambda$ suggests that resonators must be at least two orders of magnitude smaller than $\lambda$, which itself poses a challenging design issue. Fortunately, recent progress in microwave and RF metamaterials provides evidence that extremely subwavelength ($\lambda/a > 100$) negative-$\mu$ metamaterials can be manufactured~\cite{wiltshire_solymar04,freire_marques06,wiltshire07,freire_lapine10,kurter_anlage10}.

In studying magnetic surface resonances of metamaterial bodies, we cannot avoid dealing with the anisotropy of magnetic permeability.
Magnetic response in metamaterials is created by resonant elements, which are more likely than not to respond only to one field polarization.
Thus, from the design perspective anisotropic metamaterials~\cite{schurig_smith06} are much easier to achieve than their isotropic counterparts~\cite{wang_zhang11,wang_zhang_apl11}.
Additionally, magnetic resonators can be very thin, planar elements providing strong magnetic response only in the direction normal to their plane. A well-known example of such metamaterial elements are Split Ring Resonators~\cite{schurig_smith06}.

Thin resonators can be stacked with high number density, providing stronger magnetic response for the select orientation of the magnetic field.
The figure of merit for many applications including superlens-based imaging~\cite{wiltshire_hajnal06,freire_marques06,freire_lapine10} and wireless power~\cite{urzhumov_smith11,wang_zhang_apl11} is the inverse magnetic loss tangent at the frequency where $\Re\mu(\omega)=-1$.
Assuming the magnetic dispersion model~\cite{pendry_stewart99}
\be
\mu(\omega) = 1 - \frac{F \omega^2}{\omega(\omega + i \gamma) - \omega_{res}^2},
\ee
the aforementioned figure of merit is approximately
\be
\left| \frac {\Re \mu}{\Im \mu} \right| \approx \frac{1}{2} F Q,
\ee
where $Q=\omega_{res}/(2\gamma)\gg 1$ and $F\ll 1$. The magnetic oscillator strength $F<1$ is proportional to the oscillator number density; thus, densely stacked thin resonators can reduce the loss tangent in metamaterials.

By considering metamaterials with single-polarization response, we are naturally led to the consideration of magnetostatic (and by duality, electrostatic) resonances in indefinite~\cite{smith_schurig03} media. Electromagnetic waves in indefinite media have a hyperbolic dispersion relation~\cite{smith_schurig03}, leading to a variety of interesting phenomena, including photonic van Hove singularities~\cite{urzh_shvets_pre05} and extremely large photonic density of states~\cite{jacob_shalaev10}. Since indefinite media have an unusually high density of states, applying the known eigenmode methods to them may be difficult, as discussed in the subsequent section. Already in driven EM problems, indefinite media can present numerical challenges, such as ill-conditioned stiffness matrices arising in the Finite Element Method. It is therefore both surprising and practically useful that indefinite-metamaterial bodies support isolated dipole resonances that couple strongly to uniform magnetic fields. In the final section of this paper, we show how these resonances can be utilized in magnetic levitation applications.

\section{\label{sec:shape_resonance} Magnetostatic surface resonances in bodies with negative or indefinite magnetic permeability}

\begin{figure}
\centering
\begin{tabular}{c}
\includegraphics[width=0.6\columnwidth]{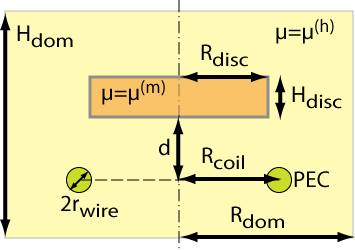}
\end{tabular}
\caption{(color online).
Schematic of the cylindrically-symmetric structure and the simulation domain used for calculations magnetostatic surface resonances.
The PEC wire domain is included only in the magnetic force calculations described in Sec.~\ref{sec:force}.
}
\label{fig:geometry}
\end{figure}

In exact analogy with electrostatic resonances, magnetostatic surface resonances can be obtained
as source-free solutions of the equation
\be \vec \nabla \mu \vec \nabla \phi^{(m)} = 0,
\ee
where $\phi^{(m)}(\vec r)$ is the magnetostatic potential. In what follows, the superscript $(m)$ is omitted. For the particular case of a homogeneous, possibly anisotropic, magnetic object with permeability $\mu^{(m)}(\omega)$ occupying a finite domain $\Omega_m$ embedded in a homogeneous host medium with a permeability tensor $\mu^{(h)}(\omega)$, permeability can be expressed through the $\theta$-function of the domain $\Omega_m$:
\be \mu(\vec x, \omega) = \mu^{(h)}(\omega) + \left(\mu^{(m)}(\omega)-\mu^{(h)}(\omega)\right)\theta(\vec x),
\ee
where by definition $\theta(\vec x)$ equals 1 if $\vec x \in\Omega_m$, and zero otherwise.

For simplicity, we assume here that $\mu^{(h)}\equiv \mu_h$ is isotropic, and $\mu^{(m)}=\diag(\mu_x,\mu_y,\mu_z)$. In metamaterials, one can control all three principal values of $\mu^{(m)}(\omega)$ individually; thus, the following Sturm-Liouville (eigenvalue) problem can be postulated:
\be \mu_h \nabla (1-\theta) \nabla \phi + (\mu_x \d_x  \theta \d_x + \mu_y \d_y  \theta \d_y)\phi = -\lambda \d_z  \theta \d_z \phi,
\label{eq:phim_aniso}
\ee
where $\lambda\equiv \mu_z$ is the unknown eigenvalue and $\mu_{x,y}$ are constants.
From now on, we assume that the magnetic object is uniaxially anisotropic ($\mu_x=\mu_y$), in which case Eq.~(\ref{eq:phim_aniso}) simplifies to
\be \mu_h \vec \nabla (1-\theta) \vec \nabla \phi + \mu_x \vec \nabla_{xy} \theta \vec \nabla_{xy} \phi = -\lambda \d_z  \theta \d_z \phi,
\label{eq:phim_uniax}
\ee
where $\vec \nabla_{xy}\equiv \hat x \d_x + \hat y \d_y$.
Alternatively, one may postulate an eigenvalue problem,
\be \mu_h \vec \nabla (1-\theta) \vec \nabla \phi + \mu_z \d_z  \theta \d_z \phi = - \lambda \vec \nabla_{xy} \theta \vec \nabla_{xy} \phi,
\label{eq:phim_uniax_2}
\ee
in which $\mu_x\equiv\mu_y=\lambda$ is treated as the eigenvalue. Finally, if the medium is isotropic, postulating $\mu_x\equiv\mu_y\equiv\mu_z=\lambda$ results in the following eigenvalue problem:
\be \mu_h \vec \nabla (\theta-1)\vec\nabla \phi = \lambda \vec \nabla \theta \vec \nabla \phi.
\label{eq:phim_iso}
\ee

Equations~(\ref{eq:phim_aniso},\ref{eq:phim_uniax},\ref{eq:phim_uniax_2},\ref{eq:phim_iso}) belong to the class of generalized eigenvalue problems (GEP), that is, they can be written as $A \phi = \lambda B \phi$, where $A$ and $B$ are linear operators; the particular case with $B=1$ is known as the standard eigenvalue problem. Analytical properties of GEP can be improved by a rational transformation of the eigenvalue, $\lambda=(a\lambda'+b)/(\lambda'+c)$, which leads to a new GEP $A'\phi = \lambda' B' \phi$,
with the new linear operators $A'=A c - B b$ and $B'=a B-A$. With a proper choice of $a,b,c$
one can sometimes make the new operators $A'$ and $B'$ positive definite, and also confine
all eigenvalues to a finite interval such as $(0, 1)$. For instance, transformation $\lambda'=1/(1-\mu_z/\mu_h)$  converts Eq.~(\ref{eq:phim_iso}) to the canonical form postulated by Bergman et al.~\cite{bergman_stroud92}:
\be
\vec \nabla \theta \vec \nabla \phi = \lambda' \vec \nabla^2 \phi.
\label{eq:phim_iso_transformed}
\ee

The eigenvalues of the problem (\ref{eq:phim_iso_transformed}) are known~\cite{bergman_stroud92} to be real and confined on the interval $(0, 1)$, that is, in the region(s) with $\Re [\mu_z(\omega)/\mu_h(\omega)]<0$. In what follows, we assume $\mu_h=1$. The eigenvalues can be found analytically only for several geometries that possess a high degree of symmetry, such as circular and elliptic cylinders~\cite{fredkin03}, spheres~\cite{sihvola94}, hemispheres~\cite{sihvola08} and so on. For most other shapes, the problem (\ref{eq:phim_iso_transformed}) must be solved numerically. However, once the eigenmodes of Eq.~(\ref{eq:phim_iso_transformed})) are found, analytically or numerically, the response of the structure with a complex permeability $\mu_z\equiv\mu'_z+i\mu''_z$ to a uniform magnetic field $H_0$ can always be written in the following form, no matter how complicated the shape of the structure:
\be
\phi(\vec r) = \phi_0(\vec r) +\sum_n \frac{\lambda'_n}{\lambda'-\lambda'_n}\frac{(\phi_n,\phi_0)}{(\phi_n,\phi_n)}\phi_n(\vec r),
\label{eq:Bergman_driven_sol}
\ee
where the complex permeability enters as $\lambda'=1/(1-\mu_z/\mu_h)$, the functions $\phi_n(\vec r)$ are the eigenmodes corresponding to the eigenvalues $\lambda'_n$ (each labeled with an integer subscript $n$), and $\phi_0=-H_0 z$ is the magnetostatic potential of the uniform magnetic field polarized in the $z$-direction. The scalar product in (\ref{eq:Bergman_driven_sol}) is defined~\cite{urzh_shvets_ssc08} as $(\phi,\psi)=\int_\Omega \theta \vec \nabla \phi^* \cdot\vec \nabla \psi$. Additionally, as shown in Ref.~\cite{urzh_shvets_ssc08}, the total (magnetic) dipole moment of the inclusion $\Omega_m$ is always given by the expression
\be
m_z = H_0 V \sum_n \frac{f_n}{\lambda'_n-\lambda'},
\label{eq:Bergman_pol}
\ee
no matter how complicated the shape of $\Omega_m$ is; general expressions for $f_n$ in terms of the eigenmodes of each resonance can be found in Ref.~\cite{urzh_shvets_ssc08}. Thus, calculation of a magnetostatic response of a given body can be reduced to the calculation (analytical or numerical) of the set of numbers $\lambda'_n$ and $f_n$ that depend only upon the shape of $\Omega_m$ and characterize its magnetic polarizability completely. The numbers $\lambda'_n$, different for each unique shape, determine the  values of the magnetic permeability $\mu_z$ at which magnetostatic resonances occur, and $f_n$ determines the strength of that resonance.

A general observation regarding the shape of the magnetic response as a function of permeability of the body can be made from the equation (\ref{eq:Bergman_pol}). As a function of the complex parameter $\lambda'=1/(1-\mu_z)=\lambda'_r+i\lambda'_i$, magnetic moment $m_z$  is a sum of simple-pole terms. For any passive medium, $\Im \mu_z$ and consequently $\Im \lambda'=\lambda'_i$ is positive; as we pointed out, the eigenvalues $\lambda'_n$ are always real~\cite{bergman_stroud92}.
If we assume a constant value of $\lambda'_i>0$ and plot the real and imaginary parts of one resonant term from Eq.~\ref{eq:Bergman_pol},
$\frac{1}{\lambda'_n-\lambda'}$
as a function of the real parameter $\lambda'_r$, we trivially obtain
\ba
\Re \frac{1}{\lambda'_n-\lambda'} = \frac{\lambda'_n-\lambda'_r}{(\lambda'_n-\lambda'_r)^2 + (\lambda'_i)^2}, \\
\Im \frac{1}{\lambda'_n-\lambda'} = \frac{\lambda'_i}{(\lambda'_n-\lambda'_r)^2 + (\lambda'_i)^2}.
\ea
Comparing these expressions with the well-known shape of resonant response functions in the frequency domain, i.e.,
\ba
\Re \frac{1}{\omega^2_n-\omega^2-i\gamma\omega} = \frac{\omega^2_n-\omega^2}{(\omega^2_n-\omega)^2 + (\gamma\omega)^2}, \\
\Im \frac{1}{\omega^2_n-\omega^2-i\gamma\omega} = \frac{\gamma \omega}{(\omega^2_n-\omega)^2 + (\gamma\omega)^2},
\ea
we must conclude that $m_z$ as a function of $\lambda'_r$ looks like the familiar Lorentz-shaped resonance. As a function of $\mu'_z$, however, the magnetic polarizability deviates from a perfect Lorentzian shape, because $\lambda'_r \equiv (1-\mu'_z)/[(1-\mu'_z)^2+(\mu''_z)^2]$ is not a linear function of $\mu'_z$.

For an arbitrary-shape particle $\Omega_m$, the aforementioned eigenvalue problem must be solved numerically.
The GEP (\ref{eq:phim_iso_transformed}) describing the isotropic medium case was solved in Refs.~\cite{shvets_urzh_prl04,urzh_jopa07,urzh_shvets_spie07} using the Finite Element Method. We have generalized the method of Ref.~\cite{shvets_urzh_prl04} to include the anisotropic medium equations (\ref{eq:phim_aniso},\ref{eq:phim_uniax},\ref{eq:phim_uniax_2}). Numerically, the obtained eigenvalue problem produces reasonable results whenever the tensor $\mu^{(m)}(\omega)$ is negative-definite, i.e. when $\mu_x\equiv\mu_y<0$ and $\mu_z<0$; in that regime, both Eqns.~(\ref{eq:phim_uniax},\ref{eq:phim_uniax_2}) returned a sequence of well-separated eigenmodes. However, in the most interesting, {\it indefinite medium} regimes, including $\mu_x\equiv\mu_y<0, \mu_z>0$ and $\mu_x\equiv\mu_y>0, \mu_z<0$, both Eqns.~(\ref{eq:phim_uniax},\ref{eq:phim_uniax_2}) produce an extremely dense set of eigenmodes, all of which have a very small dipole moment, and correspondingly negligible coupling to uniform magnetic field. Decreasing the mesh size in the FEM discretization results in even denser sets of eigenmodes; thus, no $h$-convergence can be observed.

This numerical problem stems from the well-known fact that indefinite media support waves with a hyperbolic dispersion relation~\cite{smith_schurig03}. A lossless indefinite medium thus allows wave propagation with arbitrarily large wavenumbers~\cite{smith_schurig03}; in practice, there is still a maximum (cut off) wavenumber $k_{max}$, which is related to metamaterial granularity and loss, but is typically much larger than the free-space wavenumber $k_0$. In numerical FEM solutions, $k_{max}$ is set by the mesh size $h$; shrinking the latter leads to an even larger number of eigenmodes corresponding to high $k$-waves propagating in the indefinite medium. Since no $h$-convergence can be obtained, there is no straightforward generalization of the method from Refs.~\cite{shvets_urzh_prl04,urzh_jopa07,urzh_shvets_spie07} to the indefinite medium case.

\begin{figure}
\centering
\begin{tabular}{cc}
\includegraphics[width=0.45\columnwidth]{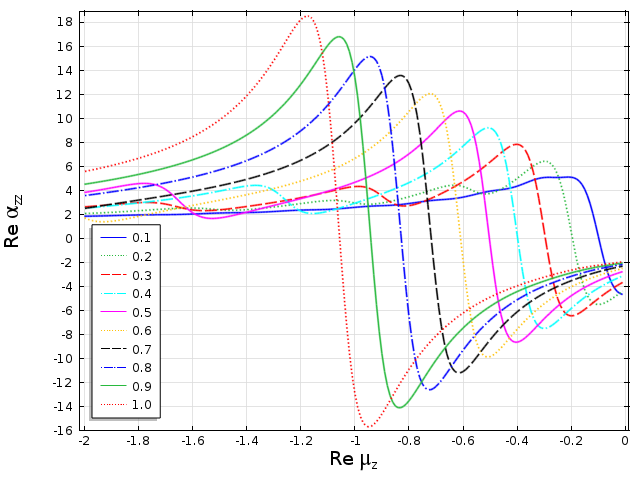}&
\includegraphics[width=0.45\columnwidth]{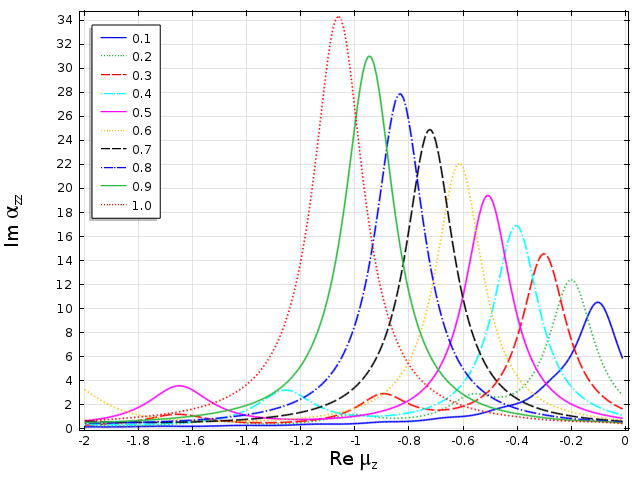}\\
(a)&(b)\\
\end{tabular}
\caption{(color online).
Real (a) and imaginary (b) part of the volume-normalized magnetic polarizability $\tilde\alpha_{zz}^{(m)}$ of a metamaterial disc,
versus the value of $\Re \mu_z$ in the disc, for various shape aspect ratios ($H/R$, indicated by numbers in the legend).
Peaks of $\Im \tilde\alpha_{zz}^{(m)}$ correspond to MSRs of the disc with non-vanishing dipolar strength.
The disc is homogeneous with permeability $\mu=\diag(1+0.1i,1+0.1i,\Re \mu_z+0.1i)$.
}
\label{fig:mu_vs_shape}
\end{figure}

On the other hand, it is known that at least simple shape objects filled with indefinite medium support well-defined dipole resonances, which couple strongly to the uniform field. For example, the magnetic polarizability of a homogeneous sphere with relative permeability $\mu=\diag(\mu_x,\mu_y\mu_z)$ can be found by applying EM duality to the electric polarizability derived by Sihvola~\cite{sihvola94}:
\be
\alpha^{(m)} = 3\mu_0 V\frac{(\mu_x-1)(\mu_y-1)(\mu_z-1)}{(\mu_x+2)(\mu_y+2)(\mu_z+2)},
\label{eq:sphere_aniso}
\ee
where $V$ is the volume of the sphere.
Therefore, even with $\mu_x=\mu_y>0$, there exists a strong, distinct and isolated dipole resonance at $\mu_z=-2$, situated deeply in the indefinite medium regime. Analytical solutions of the surface resonance problem are known only for a handful of simple shapes, including circular and elliptic cylinders and spheres. To the best of our knowledge, the sphere is the only essentially three-dimensional shape of an anisotropic medium object for which analytical solutions were obtained.

In order to find the eigenmodes of equations (\ref{eq:phim_aniso},\ref{eq:phim_uniax},\ref{eq:phim_uniax_2}) for an arbitrary-shape indefinite metamaterial body, we resort to solving a driven magnetostatic problem.
The numerical model includes a homogeneous (effective) medium model of a metamaterial body and a finite-size simulation domain of diameter substantially larger than the metamaterial object, but still much smaller than the free-space wavelength. A constant, uniform magnetic field $\vec H_0=H_0 \hat e_z$ is applied, which corresponds to magnetostatic potential $\phi_0=-H_0 z$; here, $H_0=1$~A/m is an arbitrary intensity of the applied magnetic field.
The metamaterial permeability is then scanned through the range of values where magnetostatic resonances are expected. For brevity, we describe the solutions only to the problem (\ref{eq:phim_uniax}), in which $\mu_z$ is treated as the unknown (eigenvalue) and $\mu_x=\mu_y=1$; the other eigenvalue problems can be treated similarly. To simulate the response in the $z$-direction, we choose the exterior boundary of the simulation domain to be a circular cylinder with its axis of revolution pointing in the $z$-direction. Dirichlet boundary conditions $\phi=0$ and $\phi=-H_0 H_{dom}$ are applied on the bottom ($z=0$) and top ($z=H_{dom}$) plates; the side walls are modeled as Neumann boundaries ($\vec n \cdot \vec\nabla\phi=0$), which implies that the magnetic field is parallel to those boundaries.

For the composite structure consisting of the metamaterial body and an homogeneous isotropic medium surrounding the metamaterial, one can introduce the relative effective permeability in the direction of the applied magnetic field:
\be
\mu_z^{\eff} = \frac{1}{A_{dom}\mu_0 H_0} \int \vec n \cdot \vec B dA,
\label{eq:mu_eff}
\ee
where $A_{dom}=\pi R_{dom}^2$ is the area of the bottom boundary of the cylindrical domain, and integration is carried over the latter.
The simulation domain for this type of calculation is depicted in Fig.~\ref{fig:geometry}.
This is the dual, magnetic version of the ``capacitor" definition of effective permittivity~\cite{urzh_shvets_spie07}. In what follows, the metamaterial is assumed to be situated in free space ($\mu_h=1$). A finite loss characterized by $\Im \mu^{(m)}>0$ is assumed in the simulations;
in the indefinite medium case, this turns out to be necessary not only for physical realism, but also to provide numerical stabilization of the resulting linear algebra problem. Without the loss, the stiffness matrix of the FEM discretization can be ill-conditioned, i.e. close to a singular matrix.

\begin{figure}
\centering
\begin{tabular}{cc}
\includegraphics[width=0.45\columnwidth]{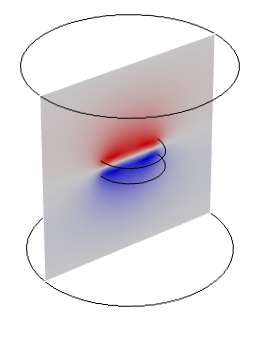}&
\includegraphics[width=0.45\columnwidth]{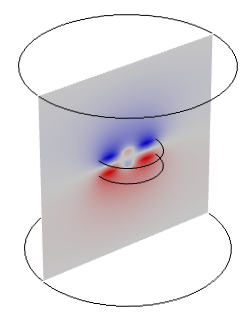}\\
(a)&(b)\\
\end{tabular}
\caption{(color online).
Magnetostatic resonances of a homogeneous disc from Fig.~\ref{fig:mu_vs_shape} and aspect ratio $H_{disc}/R_{disc}=0.5$.
Cross-sections show magnetostatic potential $\phi^{sc}=\phi-\phi_0$, calculated from Laplace equation with excitation potential $\phi_0=-z$, corresponding to uniform magnetic field of unit magnitude in the $z$-direction.
(a) Potential distribution at $\Re\mu_z= -0.49$.
(b) Potential distribution at $\Re\mu_z= -1.68$.
}
\label{fig:dipole_resonances}
\end{figure}

In the magnetostatic limit, one can readily relate the effective magnetic permeability as defined by Eq.~\ref{eq:mu_eff}, to the magnetic polarizability of the magnetizable body inside the simulation domain. Integration of the local magnetization density, $\vec M\equiv \vec B/\mu_0 - \vec H$, over the simulation domain $\Omega$, yields the total magnetic moment of the body, since it is situated in free space where $\vec M=0$:
\be
\int_\Omega \vec M dV = \int_{\Omega_m} \vec M dV \equiv \vec m.
\label{eq:int_M}
\ee
One can then introduce effective susceptibility averaged over the domain $\Omega$ in accordance with the following definition:
\be
\chi_z^\eff \equiv \frac{\int_\Omega M_z dV}{H_0 V} \equiv \frac{m_z}{H_0 V} \equiv \frac{\alpha_{zz}^{(m)}}{V},
\label{eq:chi_eff_def}
\ee
where we have also introduced the magnetic polarizability, $\alpha_{zz}^{(m)}\equiv m_z/H_0$, of the magnetizable body represented by the domain $\Omega_m\in\Omega$. In terms of dimensionless quantities -- the filling fraction $f_m=V_m/V$ and volume-normalized polarizability
$\tilde \alpha_{zz}^{(m)}\equiv \alpha_{zz}^{(m)}/V_m$, where $V_m$ is the volume of domain $\Omega_m$ -- Eq.~\ref{eq:chi_eff_def} can also be stated as $\chi_z^\eff = f_m  \tilde \alpha_{zz}^{(m)}$.

It was shown by Urzhumov et al. using the terminology of electrostatics~\cite{urzh_shvets_spie07} that, in the quasistatic limit, the quantities defined by Eq.~(\ref{eq:mu_eff},\ref{eq:chi_eff_def}) are related by the familiar equation
\be
\mu_z^\eff - 1 = \chi_z^\eff,
\ee
and consequently, $\mu_z^\eff - 1 = f_m \tilde \alpha_{zz}^{(m)}$. Thus, the calculation of $\mu_z^\eff-1$ by virtue of Eq.~\ref{eq:mu_eff}, also gives the magnetic polarizability of the body, which are directly proportional to each other.

The resonances are identified as Lorentz resonances in $\Re \mu_{\eff}$ and $\Im \mu_{\eff}$ plotted as a function of $\Re \mu_z$ (see Fig.~\ref{fig:mu_vs_shape}). The metamaterial permeability is assumed to be $\mu=\diag(1+0.1i,1+0.1i,\Re\mu_z+0.1i)$, with variable $\Re\mu_z$.
The finite metamaterial loss lets us locate and characterize the resonances that couple to uniform magnetic field by plotting $\Im \mu_{\eff}$ versus $\Re \mu_z$; such a curve features a series of well-defined absorption bands.
The abscissa of each absorption peak defines the value of $\Re \mu_z(\omega)$ (and, given the metamaterial dispersion curve $\mu_z(\omega)$, also the frequency) at which the resonance occurs; the ordinate measures the strength of coupling of the resonance to the uniform magnetic field. Fig.~\ref{fig:mu_vs_shape} shows the results for a disc-shaped metamaterial body shown schematically in Fig.~\ref{fig:geometry}; its cylindrical symmetry allows us to speed up the calculations by using the axisymmetric formulation of the magnetostatics problem in COMSOL Multiphysics~\cite{comsol}.

In the range  $-2<\Re \mu_z<0$, we observe one or two magnetic-dipole resonances of the disc, which occur at different values of $\Re\mu_z$ depending on the aspect ratio $H_{disc}/R_{disc}$, where $H_{disc}$ and $R_{disc}$ are the height and radius of the metamaterial disc, respectively. In the entire range $-\infty<\Re \mu_z<0$, there is an infinite series of such resonances occurring at progressively higher values of $-\Re\mu_z$; however only the lowest two resonances are strong enough to have practical applications. In particular, we do not find strong resonances associated with the edge of the disc; in general, sharp corners tend to produce resonances~\cite{wallen_sihvola08} with fields confined around the tip of the corner. In the case of a disc-shaped object, which has only edges and no sharp corners, such eigenmodes do not contribute significantly to the quasistatic response.

For the aspect ratio $H_{disc}/R_{disc}=0.5$, the first and the strongest resonance is obtained at $\Re\mu_z\approx-0.5$, and the second resonance --- at $\Re\mu_z\approx-1.65$. The magnetic potential distributions corresponding to these resonances are shown in Fig.~\ref{fig:dipole_resonances}. The implications of these resonances to magnetic force enhancements are discussed in the next section.

\section{\label{sec:force} Magnetic levitation of deeply subwavelength metamaterial objects}

\begin{figure}
\centering
\begin{tabular}{cc}
\includegraphics[width=0.45\columnwidth]{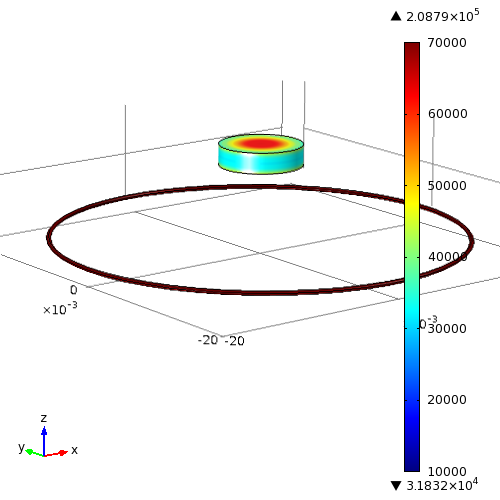}
&\includegraphics[width=0.45\columnwidth]{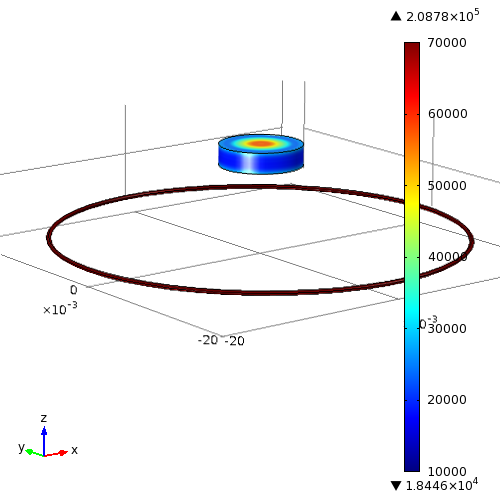}\\
(a)&(b)\\
\end{tabular}
\caption{(color online). Magnetic field ($|\vec H|$) distribution on the surface of a metamaterial disc with permeability
$\mu=\diag(1+0.1i,1+0.1i,\Re \mu_z+0.1i)$. Disc aspect ratio $H_{disc}/R_{disc}=0.5$.
(a) $\Re \mu_z=-0.5$, (b) $\Re \mu_z=-1.65$.
}
\label{fig:disc_levitation}
\end{figure}

The potential energy of a magnetic body possessing a permanent magnetic dipole moment $\vec m$,
magnetic quadrupole moment ${Q}^{(m)}$ and other multipole moments can be written as~\cite{landau_fields75}
\be
U(\vec x)=-m_j H_j-\frac{1}{6} Q^{(m)}_{jk} \frac{\d H_k}{\d x_j} - \dots;
\label{eq:potential_energy}
\ee
the magnetic field is evaluated at the geometric center of the body.
This multipolar expansion is valid whenever the body is substantially smaller than the spatial scale of magnetic field inhomogeneity
$l_{inh}=\frac{|\vec H|}{||\nabla_i H_j||}$.
Using the principle of virtual displacement, one derives the force
\be
F_i = - \frac{\d U}{\d x_i} = m_j \frac{\d H_j}{\d x_i} + \frac{1}{6} Q^{(m)}_{jk} \frac{\d H_k}{\d x_i \d x_j} + \dots
\label{eq:force}
\ee

For bodies much smaller than $l_{inh}$, the first, dipolar term is typically the dominant contribution both in the potential energy (\ref{eq:potential_energy}) and in the magnetostatic force (\ref{eq:force}), and higher order multipoles can be neglected. Assuming the geometry shown in Fig.~\ref{fig:disc_levitation}, that is, a disc of radius $R_{disc}$ over a circular coil or radius $R_{coil}$, the scale $l_{inh}$ is on the order of $R_{coil}$. For $R_{disc}\sim R_{coil}$, quadrupolar contributions can be noticeable;
to simplify our analysis, we assume $H_{disc}\le R_{disc}\ll R_{coil}$, which allows us to focus on the lowest-order term in the expansion (\ref{eq:force}), which is proportional to the magnetic dipole moment.

When the magnetic body has no permanent and only induced magnetic moment, one introduces the magnetic dipole polarizability tensor $\alpha^{(m)}_{jk}$ such that $m_j=\alpha^{(m)}_{jk} H_k$. For induced moments, the potential energy expression (\ref{eq:potential_energy})
needs to be corrected by an additional factor of $1/2$, however the force expression (\ref{eq:force}) is still correct, and it reads, in terms of the polarizability tensor,
\be
F_i =  \alpha^{(m)}_{jk} H_k \frac{\d H_j}{\d x_i} + \dots,
\label{eq:force_polarizability}
\ee
where the dots represent higher-order terms that are small relative to the dominant, dipole term for bodies of diameter $\ll l_{inh}$.
Although our main interest is to anisotropic bodies, we should remind the reader that
for isotropic bodies (i.e., belonging to an isometric symmetry class and filled with isotropic permeability), magnetic polarizability
$\alpha^{(m)}_{jk}=\alpha^{(m)} \delta_{jk}$ is a spherical tensor; in that case, Eq.~(\ref{eq:force_polarizability}) simplifies to the familiar expression
$\vec F=\frac{1}{2}\alpha^{(m)}\vec \nabla(H^2)$.
Note that this force vanishes in a uniform magnetic field, regardless of the properties of the magnetizable object.

Before we consider magnetic levitation of metamaterial bodies, let us make a crude estimate of the current in the coil required to levitate an object of an arbitrary, unoptimized shape, filled with isotropic, positive permeability. Magnetic polarizability of a sphere of volume $V$ and isotropic relative permeability $\mu_r$ is, according to Eq.~(\ref{eq:sphere_aniso}),
\be
\alpha^{(m)} = \mu_0 V \frac{3(\mu_r-1)}{\mu_r+2} \sim \mu_0 V,
\label{eq:alpha_m_iso}
\ee
where in the latter estimate we assume $\mu_r > 1$.
Magnetic induction on the axis of the coil carrying current $I$ is $B_z= \mu_0 R_{coil}^2 I/(2 d^3)$, where $d$ is the distance from the coil center. Assuming $d \sim R_{coil}$, we estimate the B-field gradient to be $\sim B/R_{coil}$, which gives the magnetic force density
\be
\frac{F_m}{V} \sim \mu_0 \frac{I^2}{R_{coil}^3} \sim 10^{-6} \frac{N}{m^3} \left( \frac{I}{1 A}\right)^2 \left( \frac{1 m}{R_{coil}}\right)^3
\ee
in SI units. Dividing this by the gravitational force density $F_g/V=\rho g$, where $g\approx 10$~m/s$^2$,
we obtain the figure of merit for magnetic levitation:
\be
\frac{F_m}{F_g} \sim 10^{-4} \left( \frac{1 g/cm^3}{\rho}\right) \left( \frac{I}{1 A}\right)^2 \left( \frac{1 cm}{R_{coil}}\right)^3.
\label{eq:levitation_FoM}
\ee
Thus, levitation of an object of density 1 g/cm$^3$ and size 1~cm or smaller using a single-turn coil of radius 1~cm requires a current of order 100 A, well beyond the practically achievable limits. It is therefore important to find a way to enhance the magnetic force by optimizing the properties of the levitating object.

From the previous section we already know that tuning the permeability of an object to one of its magnetic surface resonances results in resonant enhancement of the magnetic polarizability, and consequently, one expects a similar increase in the magnetic force. In a finite-loss medium, the enhancement is proportional to the factor $1/\Im \mu^\eff$, which arises from the denominator of Eq.~(\ref{eq:alpha_m_iso}) in which $\Re\mu^\eff$ is matched to the resonance condition. Practically, this means that a resonantly-enhanced levitation system needs a smaller current magnitude --- smaller by a factor of $|\Im \mu^\eff|^{-1/2}$ --- to achieve the same levitating force that would exist in off resonance conditions. Alternatively, with a given, attainable current magnitude, one could levitate an object heavier by the factor of $1/\Im \mu^\eff$. Loss tangents of order 0.1 are attainable~\cite{wang_zhang_apl11,urzhumov_smith11} in negative-permeability metamaterials at 10~MHz; thus metamaterials could potentially increase magnetic levitation strength by an order of magnitude.

To quantify this phenomenon more robustly, we carry out full-wave EM simulations based on the Finite Element Method. We solve the Helmholtz equation for the vector potential $\vec A$, which is linked to the magnetic induction in the usual fashion, $\vec B=\vec\nabla\times\vec A$. Again, we assume a disc-shaped metamaterial object of radius $R_{disc}$ and height $H_{disc}$ levitating above a circular coil of radius $R_{coil}$ and wire radius $r_{wire}=0.01 R_{coil}$. The simulation domain is shown schematically in Fig.~\ref{fig:geometry}.
When the disc is situated coaxially with the coil, we can take advantage of the cylindrical symmetry of the problem and use the axisymmetric version of the EM solver as provided in COMSOL Multiphysics~\cite{comsol}.
The vertical distance between the geometric center of the coil and the center of the disc is denoted $d$.

The coil is driven by a fixed AC current source, which delivers the current $I_0=250$~A at frequency 10~MHz, corresponding to the free space wavelength $\lambda_0\approx 30$~m. The metamaterial disc is modeled as a homogeneous body with permeability $\mu=\diag(1+0.1i,1+0.1i,\Re \mu_z+0.1i)$, where $\Re\mu_z$ can vary. The total force acting in the vertical direction (along the axis of revolution) is calculated by the software using surface integration of the Maxwell stress tensor; more details on this formalism can be found e.g. in Refs.~\cite{comsol,antonoyiannakis_pendry99}. Neglecting the already vanishingly small ratio $R_{disc}/\lambda_0$, there are three dimensionless parameters describing this configuration, in addition to the disc permeability: the disc aspect ratio $H_{disc}/R_{disc}$, disc size relative to the coil $R_{disc}/R_{coil}$, and finally the coil-disc distance relative to the coil size, $d/R_{coil}$.

When the ratio $R_{disc}/R_{coil}$ is small, the effect of higher-order multipole resonances can be neglected, which simplifies the picture and allows us to make a clear identification of the dipolar MSRs. For that reason, we assume $R_{disc}/R_{coil}=0.2$ here. Note that both the magnetic force associated with dipole polarizability and the gravity force scale linearly with the volume of levitating body; therefore, choosing to work with smaller bodies neither improves nor decreases the levitation figure of merit (\ref{eq:levitation_FoM}).

\begin{figure}
\centering
\begin{tabular}{cc}
\includegraphics[width=0.45\columnwidth]{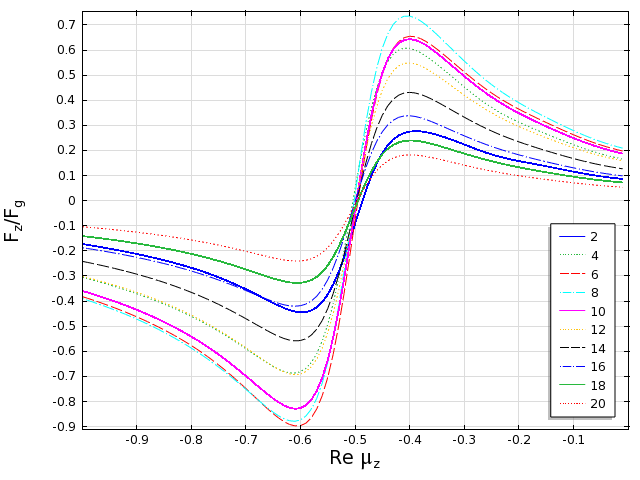}&
\includegraphics[width=0.45\columnwidth]{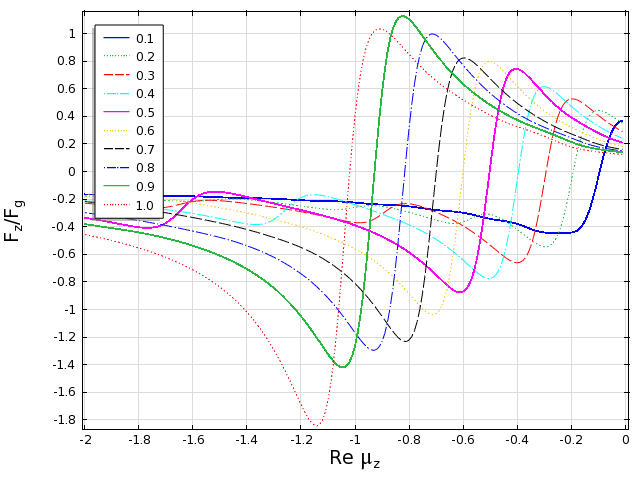}\\
(a)&(b)\\
\end{tabular}
\caption{(color online). Cycle-averaged electromagnetic force in the $z$-direction acting on a metamaterial disc with permeability
$\mu=\diag(1+0.1i,1+0.1i,\Re \mu_z+0.1i)$, versus the value of $\Re\mu_z$ in the disc.
Vertical force $F_z$ is normalized to the gravitational force $F_g=V_{disc}\rho g$, where $V_{disc}=\pi R_{disc}^2 H_{disc}$, assuming $\rho=1$~g/cm$^3$ and $g=10$~m/s$^2$.
Positive force implies repulsion from the coil (lifting force).
The current in the single-turn coil is $250$~A; coil radius $R_{coil}=2$~cm and disc radius $R_{disc}=0.2 R_{coil}=4$~mm.
(a) Force variation as a function of the distance $d$ (varied in the range 2-20~mm) between the geometric centers of the disc and the coil, for the disc aspect ratio $H_{disc}/R_{disc}=0.5$.
(b) Force variation as a function of the disc aspect ratio $H_{disc}/R_{disc}=0.1-1$, for $d=0.4 R_{coil}=8$~mm, which corresponds to the maximum force enhancement in (a).
}
\label{fig:Force_vs_mu}
\end{figure}

The dependence of the accurately computed cycle-averaged total EM force on the remaining parameters, $d/R_{coil}$, $H_{disc}/R_{disc}$, and $\mu_z$, is reported in Fig.~\ref{fig:Force_vs_mu}. First, we choose the optimum disc position relative to the coil by scanning $d/R_{coil}$.
A well-defined optimum is situated at $d=0.4 R_{coil}$ for our choice of parameters, as seen from Fig.~\ref{fig:Force_vs_mu}(a). The existence of this optimum is as follows: at sufficiently large distances from the coil, the magnetic field on the axis falls off as $1/z^3$; however, the force is also proportional to the field gradient. Due to symmetry, B-field gradient vanishes in the plane of the coil (at $d=0$). Therefore, a location exists in the vicinity of the coil where $|\d(B^2)/\d z|$ has a maximum.

From Fig.~\ref{fig:Force_vs_mu}(a) we observe that, as a function of $\Re\mu_z$,
the force resonance always occurs at a fixed value of $\Re\mu_z$, regardless of the distance $d$.
This proves that this resonance is entirely due to the properties of the metamaterial disc, and has nothing to do with the magnetic field configuration around it. In Fig.~\ref{fig:Force_vs_mu}(b), we fix the coil-disc distance $d$ at its optimum value, and study the position of the force resonance as a function of the disc shape aspect ratio, $H_{disc}/R_{disc}$.

Comparing the force resonances seen in Fig.~\ref{fig:Force_vs_mu}(b) to the resonances of magnetic polarizability (i.e. $\mu_z^\eff-1$) of the metamaterial disc in Fig.~\ref{fig:mu_vs_shape}, we observe a striking similarity. Both figures reveal a Lorentz-shape resonance, which is positioned at $\Re\mu_z=-1$ for $H_{disc}/R_{disc}=1$, $\Re\mu_z=-0.5$ for $H_{disc}/R_{disc}=0.5$, and $\Re\mu_z\approx 0$ for $H_{disc}/R_{disc} \ll 1$.
The resonance positions are in precise agreement between in Fig.~\ref{fig:Force_vs_mu}(b) and Fig.~\ref{fig:mu_vs_shape}.
This enables us to identify the strongest force resonance in Fig.~\ref{fig:Force_vs_mu}(b) with the strongest dipole MSR visible in Fig.~\ref{fig:mu_vs_shape}; for the aspect ratio $H_{disc}/R_{disc}=0.5$ it is also visualized in Fig.~\ref{fig:dipole_resonances}(a).

\section{\label{sec:conclusion} Conclusions}

In conclusion, we have theoretically demonstrated the possibility of resonant force enhancement in metamaterial bodies with negative permeability. Force enhancement is linked to the magnetostatic surface resonances (MSR) which are analogous to localized surface plasmon resonances (SPR) in negative-permittivity particles. Resonances of the magnetic force are approximately Lorentz-shaped as a function of the magnetic permeability of the metamaterial object. Consequently, the magnetic force induced in negative-$\mu$ objects can be either repulsive or attractive, depending on the object shape and its magnetic permeability. The enhancement factor with respect to the positive-permeability case is given roughly by
$1/\Im \mu^\eff$, where $\mu^\eff$ is evaluated at the resonance condition $\Re\mu^\eff=\mu_{res}$. With realistic metamaterial loss tangents of $0.1$, this phenomenon could allow magnetic levitation systems to increase the mass of levitating objects by one order of magnitude while using the same current magnitude.

\section*{\label{sec:ack} Acknowledgements}

This work was supported by the Air Force Office of Scientific Research (Contract No. FA9550-09-1-0562).
The authors are thankful to Guy Lipworth (Duke University) for help with simulations of extremely subwavelength negative-permeability metamaterials. One of the authors (YU) acknowledges discussions with Magnus Olsson (COMSOL AB) relating to maglev train modeling.

%

\bibliography{metamaterials_bib}

\end{document}